\documentclass[fleqn,twoside]{article}
\usepackage{amsmath}
\usepackage{espcrc2}
\usepackage{amssymb}
\usepackage{psfrag}
\usepackage{graphicx}
\usepackage{latexsym,amsfonts}
\begin{document}

\title{Comments on ``Rates of processes with coherent production of
different particles and the GSI time anomaly''by
C. Giunti\,,\,Phys. Lett. B {\bf 665}, 92 (2008), arXiv: 0805.0431 [hep--ph]}

\author{A. N. Ivanov $^{a,b}$\thanks{E-mail: ivanov@kph.tuwien.ac.at},
      E.  L. Kryshen${^c}$\,\thanks{E--mail: E.Kryshen@gsi.de},
      M. Pitschmann${^a}$\,\thanks{E--mail:
      pitschmann@kph.tuwien.ac.at }, P. Kienle$^{b,d}$\thanks{E-mail:
      Paul.Kienle@ph.tum.de}, \\ \addressmark{$^a$Atominstitut der
      \"Osterreichischen Universit\"aten, Technische Universit\"at
      Wien, Wiedner Hauptstra\ss e 8-10, A-1040 Wien, \"Osterreich, \\
      $^b$Stefan Meyer Institut f\"ur subatomare Physik,
      \"Osterreichische Akademie der Wissenschaften, Boltzmanngasse 3,
      A-1090, Wien, \"Osterreich},\\ $^c$ Petersburg Nuclear Physics
      Institute, 188300 Gatchina, Orlova roscha 1, Russian
      Federation,\\$^d$Physik Department, Technische Universit\"at
      M\"unchen, D--85748 Garching, Germany\\ }

\date{\today}

\begin{abstract}
  We give comments on the recent paper by Giunti (Phys. Lett. B {\bf
 665}, 92 (2008), arXiv: 0805.0431 [hep--ph] ) with a critique of our
 explanation of the experimentally observed periodic time--dependence
 of the interference term in the rate of the K--shell electron capture
 decay of the H--like ions ${^{140}}{\rm Pr}^{58+}$ and ${^{142}}{\rm
 Pm}^{60+}$ as a two--neutrino--flavour mixing. We show also that this
 phenomenon cannot be explained by a coherent mixing of two states of
 a mother ion as proposed by Giunti. \\ PACS: 12.15.Ff, 13.15.+g,
 23.40.Bw, 26.65.+t
\end{abstract}

\maketitle

Recently Litvinov {\it et al.} \cite{GSI2} have observed that the
K--shell electron capture ($EC$) decay rates of the H--like ions
${^{140}}{\rm Ce}^{58+}$ or ${^{142}}{\rm Nd}^{60+}$
\begin{eqnarray}\label{label1}
&&{^{140}}{\rm Pr}^{58+} \to {^{140}}{\rm Ce}^{58+} + \nu,\nonumber\\
&&{^{142}}{\rm Pm}^{60+} \to {^{142}}{\rm Nd}^{60+} + \nu,
\end{eqnarray}
have unexpected oscillatory structure. According to the experimental
data \cite{GSI2}, the rates of the number $N^{EC}_d$ of daughter ions
${^{140}}{\rm Ce}^{58+}$ or ${^{142}}{\rm Nd}^{60+}$
\begin{eqnarray}\label{label2}
  \frac{dN^{EC}_d(t)}{dt} = \lambda^{(\rm H)}_{EC}(t)\, N_m(t),
\end{eqnarray}
where $N_m(t)$ is the number of mother H--like ions ${^{140}}{\rm
Pr}^{58+}$ or ${^{142}}{\rm Pm}^{60+}$\cite{GSI2} and $\lambda^{(\rm
H)}_{EC}(t)$ is the $EC$--decay rate, are periodic functions, caused
by a periodic time--dependence of the $EC$--decay rates
\begin{eqnarray}\label{label3}
  \lambda^{(\rm H)}_{EC}(t) = \lambda^{(\rm H)}_{EC}\,\Big(1 + a_{EC}\,
\cos\Big(\frac{2\pi t}{T_d} + \phi\Big)\Big)
\end{eqnarray}
with a period $T_d \simeq 7\,{\rm sec}$ and an amplitude $a_{EC} \simeq
0.20$.

We have proposed in \cite{Ivanov2} an explanation of the periodic
time--dependence of the $EC$--decay rates as the interference of two
massive neutrinos $\nu_1$ and $\nu_2$ with masses $m_1$ and $m_2$,
respectively. The period $T_d$ of the time--dependence has been
related to the difference $\Delta m^2_{21} = m^2_2 - m^2_1$ of the
squared neutrino masses $m_2$ and $m_1$
\begin{eqnarray}\label{label4}
\frac{2\pi}{T_d} = \frac{\Delta m^2_{21}}{2\gamma M_m},
\end{eqnarray}
where $M_m$ is the mass of the mother ion and $\gamma = 1.43$ is a
Lorentz factor \cite{GSI2}.

For the calculation of the $EC$--decay rate we have used the standard
weak interaction Hamilton operator
\begin{eqnarray}\label{label5}
H_W(t) = \int d^3x\,\sum_j U_{ej}{\cal H}^{(j)}_W(x),
\end{eqnarray}
where $U_{ej}$ are the matrix elements of the mixing matrix of massive
neutrinos and ${\cal H}^{(j)}_W(x)$ is defined by
\begin{eqnarray}\label{label6}
\hspace{-0.3in}&& {\cal H}^{(j)}_W(x) = \frac{G_F}{\sqrt{2}}V_{ud}
  [\bar{\psi}_n(x)\gamma^{\mu}(1 - g_A\gamma^5) \psi_p(x)]\nonumber\\
\hspace{-0.3in}&& \times [\bar{\psi}_{\nu_j}(x)
  \gamma_{\mu}(1 - \gamma^5)\psi_{e^-}(x)]
\end{eqnarray}
with standard notations \cite{Ivanov2}.

The amplitude $A(m \to d + \nu)$ of the $EC$--decay $m \to d + \nu$,
where $m$, $d$ and $\nu$ are the mother ion, the daughter ion and a
neutrino, has been defined as follows
\begin{eqnarray}\label{label7}
A(m \to d + \nu) =\sum_j U_{e j}A(m \to d + \nu_j),
\end{eqnarray}
where the coefficients $U_{ej}$ testify that the electron couples to
the electron neutrino. In turn, the amplitude $A(m \to d + \nu_j)$ is
equal to
\begin{eqnarray}\label{label8}
\hspace{-0.3in}&&A(m \to d + \nu_j) = -\int d^4x\,\langle \nu_jd|{\cal
H}^{(j)}_W(x)|m\rangle =\nonumber\\ \hspace{-0.3in}&&= - (2\pi)^4\delta^{(4)}(k_d + k_j
- k_m)\,\langle \nu_j d|{\cal H}^{(j)}_W(0)|m\rangle =\nonumber\\
\hspace{-0.3in}&&= (2\pi)^4\delta^{(4)}(k_d + k_j - k_m)\,{\cal M}(m
\to d + \nu_j)
\end{eqnarray}
with ${\cal M}(m \to d + \nu_j) = -\langle \nu_j d|{\cal
H}^{(j)}_W(0)|m\rangle$.

Recently Giunti has criticised this explanation \cite{Giunti1}.
According to Giunti \cite{Giunti1}, the correct neutrino wave function
in the final state of the $EC$--decays of the H--like ions $m \to d +
\nu$ should be taken in the form
\begin{eqnarray}\label{label9}
|\nu_e(t)\rangle = \frac{\displaystyle \sum_kA_k(t)|\nu_k\rangle}{\displaystyle 
\sqrt{\sum_j|A_j(t)|^2}}
\end{eqnarray}
with  $A_k(t)$, defined by
\begin{eqnarray}\label{label10}
A_k(t) = -i\int^t_0 d\tau\,\langle \nu_k d |H_W(\tau)|m\rangle,
\end{eqnarray}
where $H_W(\tau)$ is the weak interaction Hamilton operator
Eq.(\ref{label5}).

As has been shown in \cite{Ivanov3}, such a wave function contradicts
the principles both of standard time--dependent perturbation theory
\cite{QM1}--\cite{QM4} and of quantum field theory \cite{SS61,Bjorken}.

In the recent paper \cite{Giunti2} Giunti has undertaken a new attempt
to refute the explanation, proposed in \cite{Ivanov2}, of the
experimental data by GSI \cite{GSI2}. Below we comment on Giunti's
analysis of the $EC$--decay rate.

\subsubsection*{The $EC$--decay rate}

According to Giunti \cite{Giunti2}, the decay probability $P_{m \to d +
\nu}$, defined by
\begin{eqnarray}\label{label11}
P_{m \to d + \nu} = |A(m \to d + \nu)|^2
\end{eqnarray}
with $A(m \to d + \nu)$ given by Eq.(\ref{label7}), is not equal to
\begin{eqnarray}\label{label12}
P_{m \to d + \nu} \neq \sum_j |U_{e j}|^2 |A(m \to d + \nu_j)|^2
\end{eqnarray}
In addition Giunti claims that the decay probability $P_{m \to d +
\nu}$ Eq.(\ref{label11}) has an incorrect massless limit $m_j \to 0$,
namely
\begin{eqnarray}\label{label13}
&&P_{m \to d + \nu} = \lim_{m_j \to 0}|A(m \to d + \nu)|^2 =
\nonumber\\ && = |A(m \to d + \nu)|^2_{\rm SM}\Big|\sum_j U_{e j}\Big|^2,
\end{eqnarray}
whereas the correct limit is
\begin{eqnarray}\label{label14}
&&P_{m \to d + \nu} = \lim_{m_j \to 0}|A(m \to d + \nu)|^2 =
\nonumber\\ && = |A(m \to d + \nu)|^2_{\rm SM}\sum_j|U_{ej}|^2 = \nonumber\\
&&= |A(m \to d + \nu)|^2_{\rm SM},
\end{eqnarray}
where $|A(m \to d + \nu)|^2_{\rm SM}$, calculated in the Standard
Model of electroweak interactions of heavy ions, is equal to
\begin{eqnarray}\label{label15}
|A(m \to d + \nu)|^2_{\rm SM} = \lim_{m_j \to 0}|A(m
\to d + \nu_j)|^2.
\end{eqnarray}
The incorrectness of these assertions is clearly seen if one takes
into account correctly the contribution of the $\delta$--functions
$\delta^{(4)}(k_d + k_j - k_m)$, describing the conservation of energy
and 3--momentum in the $EC$--decays.

Substituting Eq.(\ref{label8}) into Eq.(\ref{label11}) we get
\begin{eqnarray}\label{label16}
&&P_{m \to d + \nu} = |A(m \to d + \nu)|^2 =
\nonumber\\ 
&&= \sum_j |U_{ej}|^2[(2\pi)^4 \delta^{(4)}(k_d + k_j -
  k_m)]^2\nonumber\\ \hspace{-0.3in}&&\times |{\cal M}(m \to d +
  \nu_j)|^2 + 2\sum_{i > j}{\rm Re}[U^*_{ie}U_{ej}\nonumber\\
&&\times {\cal M}^*(m\to d + \nu_i){\cal M}(m \to d +
  \nu_j)]\nonumber\\
&&\times [(2\pi)^4 \delta^{(4)}(k_d + k_i - k_m)]\nonumber\\
&&\times [(2\pi)^4 \delta^{(4)}(k_d + k_j - k_m)].
\end{eqnarray}
For subsequent calculations one has to use the relations \cite{SS61}
\begin{eqnarray}\label{label17}
&&[(2\pi)^4 \delta^{(4)}(k_d + k_j -
  k_m)]^2 =\nonumber\\ 
&& = VT\,(2\pi)^4 \delta^{(4)}(k_d + k_j -
  k_m),
\end{eqnarray}
where $VT = (2\pi)^4\delta^{(4)}(0)$ is the volume of the space--time
\cite{SS61}, and
\begin{eqnarray}\label{label18}
&& [(2\pi)^4 \delta^{(4)}(k_d + k_i - k_m)]\nonumber\\
&&\times [(2\pi)^4 \delta^{(4)}(k_d + k_j - k_m)] = 0
\end{eqnarray}
The relation Eq.(\ref{label18}) is valid, since $k_i \neq k_j$. Due to
the relations Eqs.(\ref{label17}) and (\ref{label18}) the decay
probability $P_{m\to d + \nu}$ takes the form
\begin{eqnarray}\label{label19}
\hspace{-0.3in}&&P_{m \to d + \nu} = |A(m \to d + \nu)|^2 = VT\sum_j |U_{ej}|^2
\nonumber\\
\hspace{-0.3in}&&\times\,(2\pi)^4 \delta^{(4)}(k_d + k_j -
  k_m)\,|{\cal M}(m \to d +
  \nu_j)|^2 =\nonumber\\
\hspace{-0.3in}&&= \sum_j |U_{e j}|^2 |A(m \to d + \nu_j)|^2,
\end{eqnarray}
Thus, energy and 3--momentum conservation lead to the decay
probability, required by Giunti \cite{Giunti2}.

It is obvious that in the limit $m_j \to 0$ the
decay probability Eq.(\ref{label19}) reduces to the form of 
Eq.(\ref{label14}).
\begin{eqnarray}\label{label20}
&&P_{m \to d + \nu} = VT(2\pi)^4 \delta^{(4)}(k_d + k_{\nu} -
k_m)\nonumber\\ &&\times\,|{\cal M}(m \to d + \nu)|^2_{\rm SM}\sum_j
|U_{ej}|^2  =\nonumber\\ 
\hspace{-0.3in}&& =  VT(2\pi)^4 \delta^{(4)}(k_d + k_{\nu} -
k_m)\nonumber\\ &&\times\,|{\cal M}(m \to d + \nu)|^2_{\rm SM},
\end{eqnarray}
where $k_{\nu} = (E_{\nu}, \vec{k})$ with $E_{\nu} =
|\vec{k}\,|$ and $|{\cal M}(m \to d + \nu)|^2_{\rm SM}$ is given
by
\begin{eqnarray}\label{label21}
&&
|{\cal M}(m \to d + \nu)|^2_{\rm SM} = \nonumber\\
&&=\lim_{m_j \to 0}|{\cal M}(m
 \to d + \nu_j)|^2.
\end{eqnarray}
The $EC$--decay constant is defined by
\begin{eqnarray}\label{label22}
\hspace{-0.3in}&&\lambda_{EC} = \frac{1}{2 M_m}\frac{1}{2F + 1}\sum_{M_F =
\pm\frac{1}{2}}\int \frac{P_{m \to d + \nu}}{VT}\Big|_{m_j =
0}\nonumber\\
\hspace{-0.3in}&&\times\,\frac{d^3k_d}{(2\pi)^3 2 E_d}\,
\frac{d^3k}{(2\pi)^3 2 E_{\nu}}.
\end{eqnarray}
Thus, the $EC$--decay constant $\lambda_{EC}$, determined by
Eq.(\ref{label22}), is equal to the $EC$--decay constant, calculated
in \cite{Ivanov1} (see also \cite{Ivanov2}) within the Standard Model
of the electroweak interactions of heavy ions.

\subsubsection*{Giunti's wave function of neutrino with lepton flavour $\ell$}

Now let us make comments on Giunti's wave function of neutrino in the
final state of the $EC$--decay. In addition to the critique, expounded
in \cite{Ivanov3}, we would like to emphasize that Giunti's wave
function of the neutrino $\nu_{\ell}$ with the lepton flavour $\ell$
depends on the initial and the final states of the reaction $I_i \to
I_f + \nu_{\ell}$ in which the neutrino $\nu_{\ell}$ is produced,
where $I_i$ and $I_f$ are not necessary one--particle states. In order
to accentuate this point we propose to rewrite the wave function
Eq.(\ref{label9}) specifying the initial and final states as
\begin{eqnarray}\label{label23}
|\nu_{\ell}(t)\rangle_{I_iI_f} = \frac{\displaystyle
  \sum_kA_k(t)_{I_iI_f}|\nu_k\rangle}{\displaystyle
  \sqrt{\sum_j|A_j(t)_{I_iI_f}|^2}}
\end{eqnarray}
with $A_k(t)_{I_iI_f}$, given by 
\begin{eqnarray}\label{label24}
A_k(t) = -i\int^t_0 d\tau\,\langle \nu_k I_f |H_W(\tau)|I_i\rangle.
\end{eqnarray}
Hence, the neutrinos $(\nu_{\ell})_{I_iI_f}$ and
$(\nu_{\ell})_{I\,'_iI\,'_f}$, produced in two different reactions
$I_i \to I_f + \nu_{\ell}$ and $I\,'_i \to I\,'_f + \nu_{\ell}$, are
two different particles. They are not stable and the probability of
the transition $(\nu_{\ell})_{I_iI_f} \longleftrightarrow
(\nu_{\ell})_{I\,'_iI\,'_f}$ is equal to
\begin{eqnarray}\label{label25}
\hspace{-0.3in}&&P(\nu^{I_iI_f}_{\ell} \longleftrightarrow
\nu^{I\,'_iI\,'_f}_{\ell}) = |{_{I\,'_iI\,'_f}}\langle
\nu_{\ell}(t)|\nu_{\ell}(t)\rangle_{I_iI_f}|^2 =\nonumber\\ 
\hspace{-0.3in}&&=
\frac{\displaystyle
|\sum_kA^*_k(t)_{I\,'_iI\,'_f}A_k(t)_{I_iI_f}|^2}{\displaystyle
\sum_j|A_j(t)_{I_iI_f}|^2 \sum_{j'}|A_{j'}(t)_{I\,'_iI\,'_f}|^2},
\end{eqnarray}
where $A_k(t)_{I_iI_f} \neq A_k(t)_{I\,'_iI\,'_f}$ by definition due
to different initial and final states of the reactions $I_i \to I_f +
(\nu_{\ell})_{I_iI_f}$ and $I\,'_i \to I\,'_f +
(\nu_{\ell})_{I\,'_iI\,'_f}$, respectively. Since the number of
initial and final states $(I_i I_f)$ of the reactions producing
neutrinos $(\nu_{\ell})_{I_iI_f}$ with a lepton flavour $\ell$ is
infinite, so, according to Giunti \cite{Giunti1,Giunti2}, there is an
infinite set of neutrinos $(\nu_{\ell})_{I_iI_f}$ with a leptonic
flavour $\ell$.
 
\subsubsection*{Giunti's explanation of ``Darmstadt oscillations''}

According to Giunti \cite{Giunti2}, the interference term in the
$EC$--decay rate $m \to d + \nu_e $ comes from the mixing of the
different mass--states of the mother ion $m$. For the wave function of
the initial state of the mother ion, which is not an eigenstate of the
mass--operator, Giunti has proposed the following expression
\begin{eqnarray}\label{label26}
|m\rangle = \cos\theta\,|m'\rangle + \sin\theta\,|m''\rangle,
\end{eqnarray}
where $|m'\rangle$ and $|m ''\rangle$ are two states of the mother ion
with masses $M_{m '}$ and $M_{m ''}$, respectively, and $\theta$ is a
mixing angle.  This means that the initial state of the mother ion is
a coherent state of two eigenstates of the mass--operator $|m'\rangle$
and $|m ''\rangle$, respectively.  The final state of the $EC$--decay
is defined by the wave function $|d,\nu_e\rangle$, where $\nu_e$ is a
massless electron neutrino. Since Giunti's calculation has no relation
to the real calculation of the $EC$--decay rate, which is needed for
the comparison with the experimental data on the rate of the number of
daughter ions \cite{Ivanov2}, below we give a calculation of the
$EC$--decay rate within Giunti's approach in detail.

According to standard time--dependent perturbation theory
\cite{QM1}--\cite{QM4}, the amplitude of the $m \to d + \nu_e$ decay
is defined by (see also Eq.(\ref{label10}) and
\cite{Giunti1,Giunti2}),
\begin{eqnarray}\label{label27}
&&A(m \to d + \nu_e)(t) =\nonumber\\ &&= -i\int^t_0d\tau\,
\langle \nu_e,d|H_W(\tau)|m\rangle,
\end{eqnarray}
where $|m\rangle$ is the wave function Eq.(\ref{label26}) and the weak
interaction Hamilton operator $H_W(t)$ takes the form \cite{Ivanov1}
\begin{eqnarray}\label{label28}
  \hspace{-0.3in}&&H_W(t) =  \frac{G_F}{\sqrt{2}}V_{ud}\!\int\! d^3x
  [\bar{\psi}_n(x)\gamma^{\mu}(1 -
  g_A\gamma^5) \psi_p(x)]\nonumber\\
\hspace{-0.3in}&&\times [\bar{\psi}_{\nu_e}(x) \gamma_{\mu}(1 -
  \gamma^5)\psi_{e^-}(x)].
\end{eqnarray}
Suppose that the wave function of the neutrino in the $EC$--decay is a
plane wave. In this case the amplitude of the $EC$--decay is equal to
\cite{QM1}--\cite{QM4,Ivanov1}
\begin{eqnarray}\label{label29}
  \hspace{-0.3in}&&A(m \to d + \nu_e)(t) =  -\,\sqrt{3}\,\sqrt{2 M_{m}
  2 E_d(\vec{q}\,) E_{\nu_e}(\vec{k}\,)}\nonumber\\
 \hspace{-0.3in}&&\times\,{\cal M}_{\rm GT}\, \langle
  \psi^{(Z)}_{1s}\rangle\,(2\pi)^3\,\delta^{(3)}(\vec{q} +
  \vec{k}\,)\,\Big[\cos\theta\,e^{\textstyle\,i\frac{\Delta
  E'}{2}t}\nonumber\\
\hspace{-0.3in}&&\times\,\frac{\sin\Big(\frac{\Delta
  E'}{2}t\Big)}{\Big(\frac{\Delta E'}{2}\Big)}+ \sin\theta\,e^{\textstyle\,i\frac{\Delta
  E''}{2}t}\,\frac{\sin\Big(\frac{\Delta
  E''}{2}t\Big)}{\Big(\frac{\Delta E''}{2}\Big)}\Big]\nonumber\\
\hspace{-0.3in}&&\times\,\delta_{M_F,
  -\frac{1}{2}},
\end{eqnarray}
where $\Delta E' = E_d(\vec{q}\,) + E_{\nu_e}(\vec{k}\,) - M_{m'}$ and
$\Delta E'' = E_d(\vec{q}\,) + E_{\nu_e}(\vec{k}\,) - M_{m''}$,
$\vec{q}$ and $\vec{k}$ are 3--momenta of the daughter nucleus and the
neutrino, $E_d(\vec{q}\,)$ and $E_{\nu_e}(\vec{k}\,)$ are the energies
of the daughter ion and the neutrino, respectively, ${\cal M}_{\rm
GT}$ is the nuclear matrix element of the Gamow--Teller transition and
$\langle \psi^{(Z)}_{1s}\rangle$ is the wave function of the bound
electron in the H--like mother ion, averaged over the nuclear density
\cite{Ivanov1}.

The rate of the neutrino spectrum is defined by \cite{Ivanov2}
\begin{eqnarray}\label{label30}
\hspace{-0.3in}&&\frac{dN_{\nu_e}(t)}{dt} = \frac{1}{2M_m}\int \frac{d^3q}{(2\pi)^3
2E_d(\vec{q}\,)}\nonumber\\ 
\hspace{-0.3in}&&\times \frac{1}{2 F + 1}\sum_{M_F =
-\frac{1}{2}}\frac{d}{dt}|A(m\to d + \nu_e)(t)|^2 =\nonumber\\
\hspace{-0.3in}&&= \frac{3}{2 F + 1}\,V\,E_{\nu_e}|{\cal M}_{\rm
  GT}|^2 |\langle
  \psi^{(Z)}_{1s}\rangle|^2\nonumber\\
\hspace{-0.3in}&&\times\,\Bigg\{2 \cos^2\theta\,\frac{\sin(\Delta
  E't)}{(\Delta E')} + 2 \sin^2\theta\,\frac{\sin(\Delta E''t)}{(\Delta
  E'')}\nonumber\\
\hspace{-0.3in}&&+ \sin 2\theta\,\Bigg[\frac{\sin\Big(\frac{\Delta
  E'}{2}t\Big)}{\Big(\frac{\Delta E'}{2}\Big)}\,\cos \Big(\frac{\Delta
  E'}{2} - \Delta E''\Big)t\Big)\nonumber\\
\hspace{-0.3in}&& + \frac{\sin\Big(\frac{\Delta
  E''}{2}t\Big)}{\Big(\frac{\Delta E''}{2}\Big)}\,\cos \Big(\frac{\Delta
  E''}{2} - \Delta E'\Big)t\Big)\Bigg]\Bigg\}
\end{eqnarray}
Here we have used the relation 
\begin{eqnarray}\label{label31}
[(2\pi)^3\,\delta^{(3)}(\vec{q} +
  \vec{k}\,)]^2 = V\,(2\pi)^3\,\delta^{(3)}(\vec{q} + \vec{k}\,),
\end{eqnarray}
  where $(2\pi)^3 \delta^{(3)}(\vec{0}\,) = V$ is the normalisation
  volume \cite{SS61}. For sufficiently long time we get
  \cite{QM1}--\cite{Bjorken}
\begin{eqnarray}\label{label32}
\hspace{-0.3in}&&\frac{dN_{\nu_e}(t)}{dt} = \frac{3}{2 F + 1}\,V\,E_{\nu_e}|{\cal M}_{\rm
  GT}|^2 |\langle
  \psi^{(Z)}_{1s}\rangle|^2\nonumber\\
\hspace{-0.3in}&&\times\,\Big\{\cos^2\theta\,(2\pi)\delta(\Delta E')
+ \sin^2\theta\,(2\pi)\,\delta(\Delta E'')\nonumber\\
\hspace{-0.3in}&&+ \sin 2\theta\,\Big[(2\pi)\,\delta(\Delta E') +
(2\pi)\,\delta(\Delta E'')\Big]\nonumber\\
\hspace{-0.3in}&&\times\,\cos (\Delta M_m t)\Big\},
\end{eqnarray}
where $\Delta M_m = M_{m'} - M_{m''}$, $\Delta E' = E_d(\vec{k}\,) +
E_{\nu_e}(\vec{k}\,) - M_{m'}$ and $\Delta E'' = E_d(\vec{k}\,) +
E_{\nu_e}(\vec{k}\,) - M_{m''}$. The $EC$--decay rate
$\lambda^{(m)}_{EC}(t)$ from the coherent state $|m\rangle$ is defined
by
\begin{eqnarray}\label{label33}
\hspace{-0.3in}&&\lambda^{(m)}_{EC}(t) = \int \frac{d^3k}{(2\pi)^3
  2E_{\nu_e}}\,\frac{1}{V}\,\frac{dN_{\nu_e}(t)}{dt}
\end{eqnarray}
Substituting Eq.(\ref{label32}) into Eq.(\ref{label33}) and
integrating over the neutrino phase volume we get
\begin{eqnarray}\label{label34}
\lambda^{(m)}_{EC}(t) = \lambda_{EC}\Big\{1 + 2 \sin 2\theta \cos
(\Delta M_m t)\Big\},
\end{eqnarray}
where $\lambda_{EC}$ is the $EC$--decay constant, calculated in
\cite{Ivanov1}.

Thus, one can show that for the initial state of the mother ion, given
by Eq.(\ref{label26}), the $EC$--decay rate has a periodic
interference term with a period $T_d$, defined by the mass difference
$\Delta M_m = M_{m'} - M_{m''}$, which is equal to
\begin{eqnarray}\label{label35}
\Delta M_m = \frac{2\pi\gamma \hbar }{T_d c^2} = 8.45\times 10^{-16}\,{\rm
eV/c^2},
\end{eqnarray}
where $\gamma = 1.43$ is the Lorentz factor \cite{GSI2}. A reduction
of the $EC$--decay rate Eq.(\ref{label34}) to the experimental shape
\cite{GSI2}
\begin{eqnarray}\label{label36}
\lambda_{EC}(t) = \lambda_{EC}\Big\{1 + a_{EC}\cos \Big(\frac{2\pi
t}{T_d} + \phi\Big)\Big\}
\end{eqnarray}
with $a_{EC} \simeq 0.20$ \cite{GSI2} can be carried out by changing
$|m'\rangle \to e^{\,i\phi'}\,|m'\rangle$ and $|m''\rangle \to
e^{\,i\phi''}\,|m''\rangle$, giving $\phi = \phi'- \phi''$, and
setting $\theta \simeq 2.87^0$.

The problem of such an explanation of the ``Darmstadt oscillations''
is as follows. If there exist two states of the mother ion
$|m'\rangle$ and $|m''\rangle$ with a mass--difference $\Delta M_m =
8.45\times 10^{-16}\,{\rm eV/c^2}$, giving the contribution to the
$EC$--decay through the coherent state $|m\rangle$, given by
Eq.(\ref{label26}), the contribution to the $EC$--decay should be also
from the coherent state $|\tilde{m}\rangle$
\begin{eqnarray}\label{label37}
|\tilde{m}\rangle = -\,\sin\theta\,|m'\rangle + \cos\theta\,|m''\rangle,
\end{eqnarray}
which is not also an eigenstate of the mass--operator and orthogonal
to the state $|m\rangle$. The coherent state $|\tilde{m}\rangle$,
given by Eq.(\ref{label37}), can be produced in the system of mother
ions on the same footing as the coherent state $|m\rangle$, given by
Eq.(\ref{label26}). Indeed, the mother ions, injected into the Storage
Ring, are produced by means of a fast projectile fragmentation with a
statistical population of the states $|m'\rangle$ and $|m''\rangle$,
which are eigenstates of the mass--operator.  Hence, the probabilities
$P_m$ and $P_{\tilde{m}}$ of the appearance of the coherent states
$|m\rangle$ and $|\tilde{m}\rangle$, related by $P_m + P_{\tilde{m}} =
1$, should be equal $P_m = P_{\tilde{m}} = \frac{1}{2}$ due to a
principle indistinguishability of these states.

The $EC$--decay rate $\lambda^{(\tilde{m})}_{EC}(t)$ of the
$EC$--decay $\tilde{m} \to d + \nu_e$ from the coherent state
$|\tilde{m}\rangle$ is equal to
\begin{eqnarray}\label{label38}
\lambda^{(\tilde{m})}_{EC}(t) = \lambda_{EC}\Big\{1 - 2 \sin 2\theta \cos (\Delta
M_m t)\Big\}.
\end{eqnarray}
The total $EC$--decay rate, caused by the $EC$--decays of the H--like
heavy ions from the states $|m\rangle$ and $|\tilde{m}\rangle$, is
defined by
\begin{eqnarray}\label{label39}
\hspace{-0.3in}&&\lambda_{EC}(t) = P_m\lambda^{(m)}_{EC}(t) +
P_{\tilde{m}}\,\lambda^{(\tilde{m})}_{EC}(t) = \nonumber\\
\hspace{-0.3in}&&= \lambda_{EC}\Big\{1 + 2 \sin 2\theta\,(P_m -
P_{\tilde{m}}) \cos (\Delta M_m t)\Big\}.
\end{eqnarray}
Since there is no physical reason for $P_m \neq P_{\tilde{m}}$,
setting $P_m = P_{\tilde{m}}$ one gets no interference terms in the
$EC$--decay rate of the H--like heavy ion in the approach proposed by
Giunti \cite{Giunti2}.

\subsubsection*{Summary}

We have shown that Giunti's critique of our approach is based,
technically, on the missing of the $\delta$--functions, responsible
for the conservation of energy and 3--momentum in the $EC$--decay and,
globally, on the misunderstanding of the standard procedure for the
calculation of the decay rates.

Giunti's wave functions for neutrinos in the final state of the
$EC$--decay of the H--like heavy ions or generally in any weak
interaction producing or absorbing neutrinos make no sense, since they
require an infinite number of neutrinos with a lepton flavour
$\ell$. These neutrinos are not stable and oscillate with a finite
probability.

As regards Giunti's explanation of the ``Darmstadt oscillations'' we
assert the following.  Apart from the existence of a superweak
interaction, leading to the mass--splitting of the H--like heavy ions
of order $O(10^{-15}\,{\rm eV/c^2})$, which is hardly possible in
reality, in Giunti's approach the total $EC$--decay rate of the
H--like ions should be defined by $EC$--decays from two coherent
states $|m\rangle$ and $|\tilde{m}\rangle$, given by
Eqs.(\ref{label26}) and (\ref{label37}), respectively. These coherent
states can appear in the system of mother ions with probabilities
$P_m$ and $P_{\tilde{m}}$, respectively, and related by $P_m +
P_{\tilde{m}} = 1$.  The equality $P_m = P_{\tilde{m}} = \frac{1}{2}$,
caused by a statistical equivalence of the coherent states $|m\rangle$
and $|\tilde{m}\rangle$ in the system of mother ions, injected into
the Storage Ring, shows the absence of the interference term in the
$EC$--decay rate
\begin{eqnarray}\label{label40}
\hspace{-0.3in}&&\lambda_{EC}(t) = P_m\lambda^{(m)}_{EC}(t) +
P_{\tilde{m}}\,\lambda^{(\tilde{m})}_{EC}(t) =  \lambda_{EC}.
\end{eqnarray}
Thus, Giunti's explanation of the ``Darmstadt oscillations'' has no
physical ground and it is erroneous by definition.

Our analysis of Giunti's explanation of the ``Darmstadt oscillations''
can be formulated more generally as ``a non--existence of periodic
time dependent interference terms in the $EC$--decay rates of the
H--like heavy ions for the coherence in the initial state of the
mother ions''. Indeed, the states $|m\rangle$ as well as the
orthogonal state $|\tilde{m}\rangle$ can be treated as coherent states
of $|m'\rangle$ and $|m''\rangle$, produced in the statistical system
of mother ions injected into the Storing Ring. Due to a statistical
equivalence of these states, the probabilities $P_m$ and
$P_{\tilde{m}}$ of the appearance of the coherent states $|m\rangle$
and $|\tilde{m}\rangle$ in the system of mother ions should be equal
$P_m = P_{\tilde{m}}$. This prohibits the appearance of the
interference term in the $EC$--decay rate (see Eq.(\ref{label39}) and
Eq.(\ref{label40})).

We thank M. Faber, T. Ericson and N. Troitskaya for fruitful
discussions. One of us (A. Ivanov) is grateful to N. Ivanov for
discussions of physics of Storage Rings.


\begin{thebibliography}{9}
\bibitem{GSI2} Yu. A. Litvinov {\it et al.} (the GSI Collaboration),
Phys. Lett. B {\bf 664}, 162 (2008), arXiv: 0801.2079 [nucl-ex].
\bibitem{Ivanov2}
A. N. Ivanov, R. Reda, P. Kienle, {\it 
On the time--modulation of the K--shell electron capture decay of
  H-like ${^{140}}{\rm Pr}^{58+}$ ions produced by neutrino--flavour
  mixing}, arXiv: 0801.2121 [nucl--th].
\bibitem{Giunti1} 
C. Giunti, {\it Comment on neutrino--mixing
interpretation of the GSI time anomaly}, arXiv: 0801.4639 [nucl--th].
\bibitem{Ivanov3}
A. N. Ivanov, R. Reda, P. Kienle, 
{\it  Reply on `Comment on neutrino-mixing interpretation of the 
GSI time anomaly' by C. Giunti, {\rm arXiv:0801.4639 [nucl--th]}}'',
arXiv: 0803.1289 [nucl--th] and references therein.
\bibitem{Giunti2} 
C. Giunti, 
Phys. Lett. B {\bf 665}, 92 (2005), arXiv: 0805.0431 [hep--ph].
\bibitem{QM1}
L. I. Schiff,
in {\it Quantum mechanics},
McGraw--Hill Book Co., Inc., New York, 1955.
\bibitem{QM2}
A. Messiah,
in {\it Quantum mechanics}, Vol. I, North--Holland Publishing Company, 
Amsterdam, 1961;\\
A. Messiah,
in 
{\it Quantum mechanics}, Vol. II, North--Holland Publishing Company, 
Amsterdam, 1962.
\bibitem{QM3}
L. D. Landau and E. M. Lifshitz,
in {\it Quantum mechanics, Non--relativistic theory}, 
Volume 3 of Course of Theoretical Physics,
Pergamon Press, New York, 1965.
\bibitem{QM4}
W. Greiner,
in {\it Quantum Mechanics, An Introduction},
Springer--Verlag, Berlin, 2001.
\bibitem{SS61} 
S. S. Schweber, in {\it An introduction to relativistic
    quantum field theory}, Row, Peterson and Co$\,\bullet\,$ Evanston,
  Ill., Elmsford, New York, 1961.
\bibitem{Bjorken}
J. D. Bjorken and S. D. Sidney,
im {\it Relativistische Quantenmechanik}, Band I, 
Bibliographisches Institut Mannheim,
1966;\\
J. D. Bjorken and S. D. Sidney,
im {\it Relativistische Quantenfeldtheorie}, Band II, 
Bibliographisches Institut Mannheim,
1967.
\bibitem{Ivanov1}
A. N. Ivanov, M. Faber, R. Reda,  P. Kienle, arXiv: 0711.3184 [nucl--th],
(to appear in Phys. Rev. C).
\end{thebibliography}
\end{document}